\documentclass[graybox]{svmult}

\usepackage{mathptmx}       % selects Times Roman as basic font
\usepackage{helvet}         % selects Helvetica as sans-serif font
\usepackage{courier}        % selects Courier as typewriter font
\usepackage{makeidx}         % allows index generation
\usepackage{graphicx}        % standard LaTeX graphics tool
\usepackage{multicol}        % used for the two-column index
\usepackage[bottom]{footmisc}% places footnotes at page bottom
\usepackage{hyperref}        %for hyperlinks
\usepackage{soul}            % for high-lighting of text
\hypersetup{colorlinks=true,urlcolor=blue}

\usepackage{newtxtext}       % 
\usepackage[varvw]{newtxmath}       % selects Times Roman as basic font

\usepackage[square,numbers]{natbib}
\makeindex             % used for the subject index
\begin{document}
%\tableofcontents{}
\title*{In-orbit background for X-ray detectors}
\author{Riccardo Campana\thanks{corresponding author}}
\institute{Riccardo Campana \at INAF/OAS-Bologna, Via Gobetti 101, I-40129 Bologna (Italy),  \\ \email{riccardo.campana@inaf.it}
%\and Second Author \at Institute 2, Address of Institute 2 \email{name2@email.address}
}
\maketitle
\abstract{
In-orbit background is an unavoidable feature of all space-borne X-ray detectors, and arises both from cosmic sources (diffuse or point-like) and from the interaction of the detectors themselves with the space environment (primary or secondary cosmic rays, geomagnetically trapped particles, activation of spacecraft structures). In this chapter the main background sources are discussed, with their principal effects on the various detector types, and simulation and mitigation strategies are described.
}

\section{Keywords} 
%Please provide keywords required to facilitate search of chapter on web; maximum 10 keywords.
X-ray detectors, Background, Space environment, CXB, Cosmic rays, Activation

\section{Introduction}
Essentially all detectors are affected by some form of \emph{noise}, which can have different origins, and will ultimately limit the sensitivity of the instrument and its performance in measuring a physical quantity (\emph{signal}). Some sources of noise are intrinsic to the detector and its operation (e.g., the electronic noise) while others depend on the environment in which the measurement is performed. In particular, for a high energy astrophysics X-ray instrument, the space environment will induce some form of \emph{background} that will affect the observations of a cosmic source.

A correct understanding of the background is of paramount importance for the accurate interpretation of the scientific measurement. Background properties will depend in turn on the architecture of the detector, on the details of the environment in which it will operate and on the type of measurement.

In this chapter we will perform an overview of the main characteristics of the in-orbit background for a X-ray detector. In Section~\ref{s:spaceenv} the various sources of background for different orbits are summarized, while in Section~\ref{s:radeffects} their impact on different types of detectors is outlined. In Section~\ref{s:montecarlo} background simulation and estimation methods will be discussed.

\section{The space environment for a X-ray mission}\label{s:spaceenv}

A X-ray detector for high energy astrophysics will unavoidably have to be operated in space, since X-rays are absorbed by the Earth's atmosphere. Different missions with different objectives will fly in different orbits, and each will be characterised by its peculiar environment, resulting from the interplay of various physical processes in the local Universe and in the solar system.

\subsection{Orbits and their characteristics}

\begin{figure}[htbp]
\centering
\includegraphics[width=0.8\textwidth]{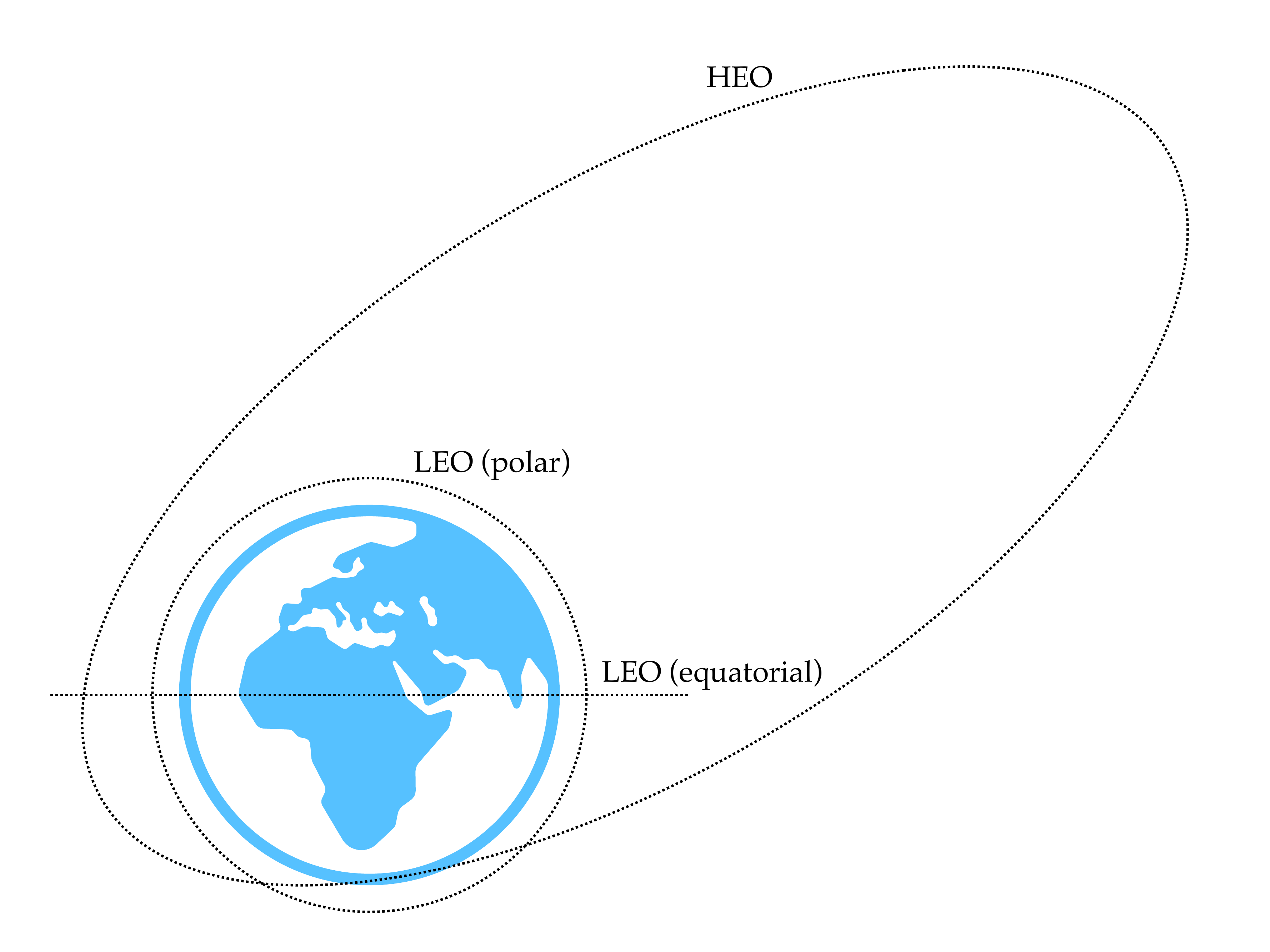}
\caption{Various types of Earth orbits.}
\label{f:orbits}
\end{figure}

\begin{figure}[htbp]
\centering
\includegraphics[width=\textwidth]{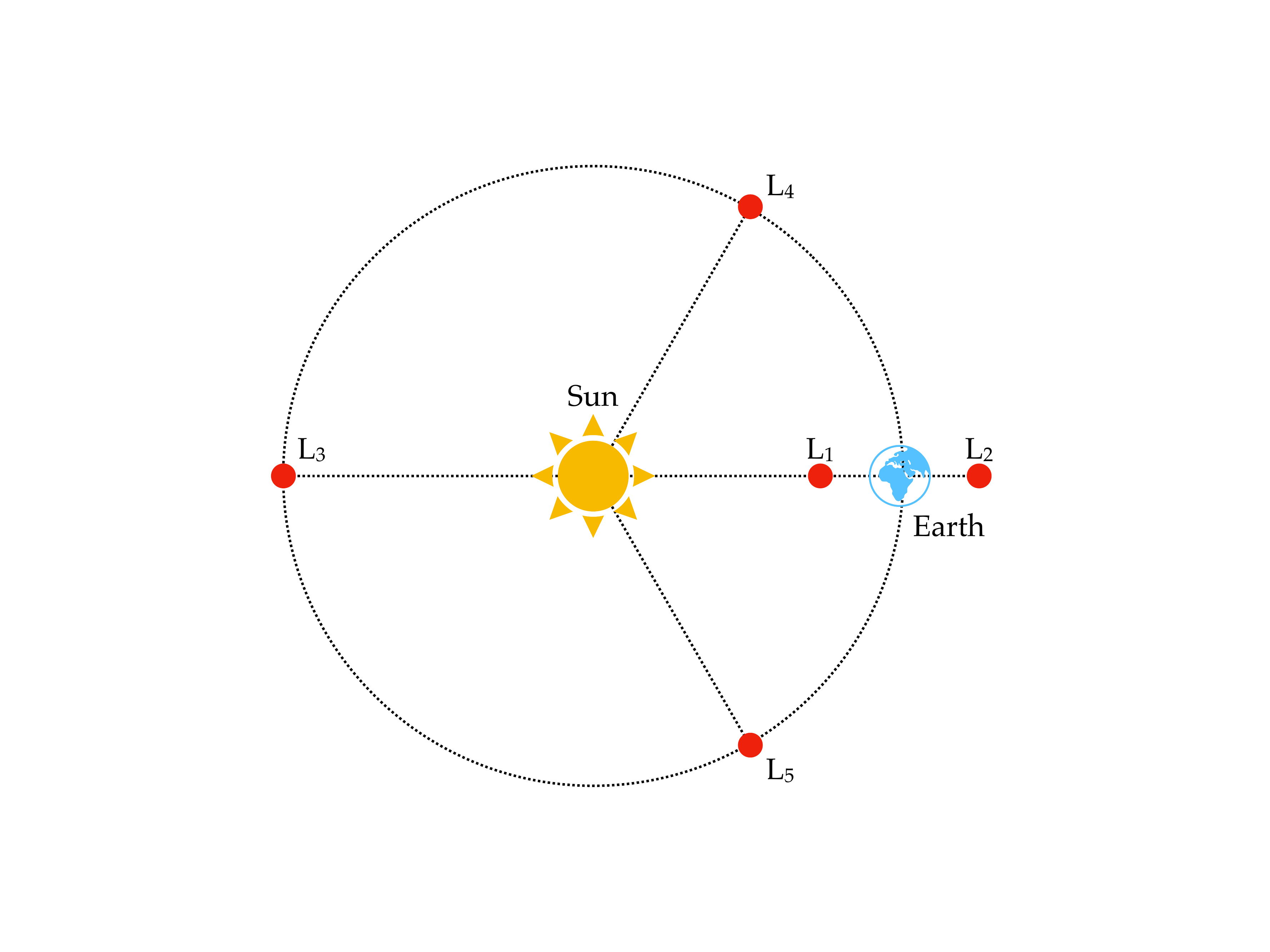}
\caption{The Lagrange points in the Sun-Earth system. The distance between $L_1$ and $L_2$ from the Earth is about 1.5 million km (0.01 AU). The other Lagrangian points ($L_3$, $L_4$ and $L_5$) are essentially not used for any space mission.}
\label{f:lagrange}
\end{figure}

The Earth orbits usually chosen for scientific satellites (at least in the field of high- energy astrophysics) can be divided in:
\begin{enumerate}
\item LEOs (\emph{Low Earth Orbits}): mostly circular orbits, below $\sim$2000~km altitude\footnote{The lower altitude limit for an orbit is given by the atmospheric braking-induced orbital decay. Satellites at altitudes below $\sim$300~km have a very short lifetime before re-entering.}. Examples: Fermi, AGILE, RXTE. Almost all human spaceflights have been flown in a LEO (Figure~\ref{f:orbits}).
\item HEOs (\emph{Highly Elliptical Orbits}, which can be thought as a subclass of High Earth Orbits): elliptical orbits, with a very high apogee ($>$36,000 km).
Examples: INTEGRAL, Chandra, XMM-Newton (Figure~\ref{f:orbits}).
\item Lagrangian Points ($L_1$, $L_2$): equilibrium points at 1.5 million km from Earth. Examples: Spectrum-Röntgen Gamma, Planck, Herschel (Figure~\ref{f:lagrange}).
\end{enumerate}

LEOs have the advantages of providing a low overall background environment, and the required launcher $\Delta v$\footnote{That is, the impulse per unit mass needed to perform the orbital insertion. In other words, a measure of the required fuel  for the launch.} is reasonably small, thus leading to higher allowable payload masses. However, the Earth occupies a significant fraction of the sky (e.g., at 600 km altitude about 30\% of the sky is blocked by the Earth) thus leading to a significant source occultation time. While a careful pointing plan may alleviate this problem, this usually gives rise to a low observation duty-cycle. An additional factor is that the Earth itself is a source of radiation (albedo). A satellite in a HEO, on the other hand, will spend almost all the time far away from the Earth, allowing long observations of the same target. However, the crossing of the radiation belts (Section~\ref{s:geofieldbelts}) on the inner magnetosphere leads to a highly variable background.  \emph{Halo} orbits around Lagrangian points allow an almost uninterrupted observation time, and a stable thermal environment, at the price of a high background due to unshielded cosmic rays and solar wind. In particular, the $L_2$ point is located on the geomagnetic field tail, which is a highly dynamic environment, leading to significant short-term background variations.

Several other Earth orbits exist, such as the geosynchronous/geostationary orbits (GEOs, 36,000 km altitude and $\sim$0$^\circ$ inclination), or medium Earth orbits (MEOs, between 2000 and 36,000 km altitude), but these are seldom used for X-ray astronomy experiments. Lunar orbits and interplanetary trajectories will have problematics similar to Lagrangian halo orbits, and will not be further considered.

LEOs can be further sub-classified in equatorial or low-inclination (equatorial) orbits, and high-inclination orbits (polar or sun-synchronous). Background characteristics will be different for these categories, given the dependence of the environment on the local geomagnetic conditions. 

\subsection{The geomagnetic field and the radiation belts}\label{s:geofieldbelts}
\begin{figure}[htbp]
\centering
\includegraphics[width=\textwidth]{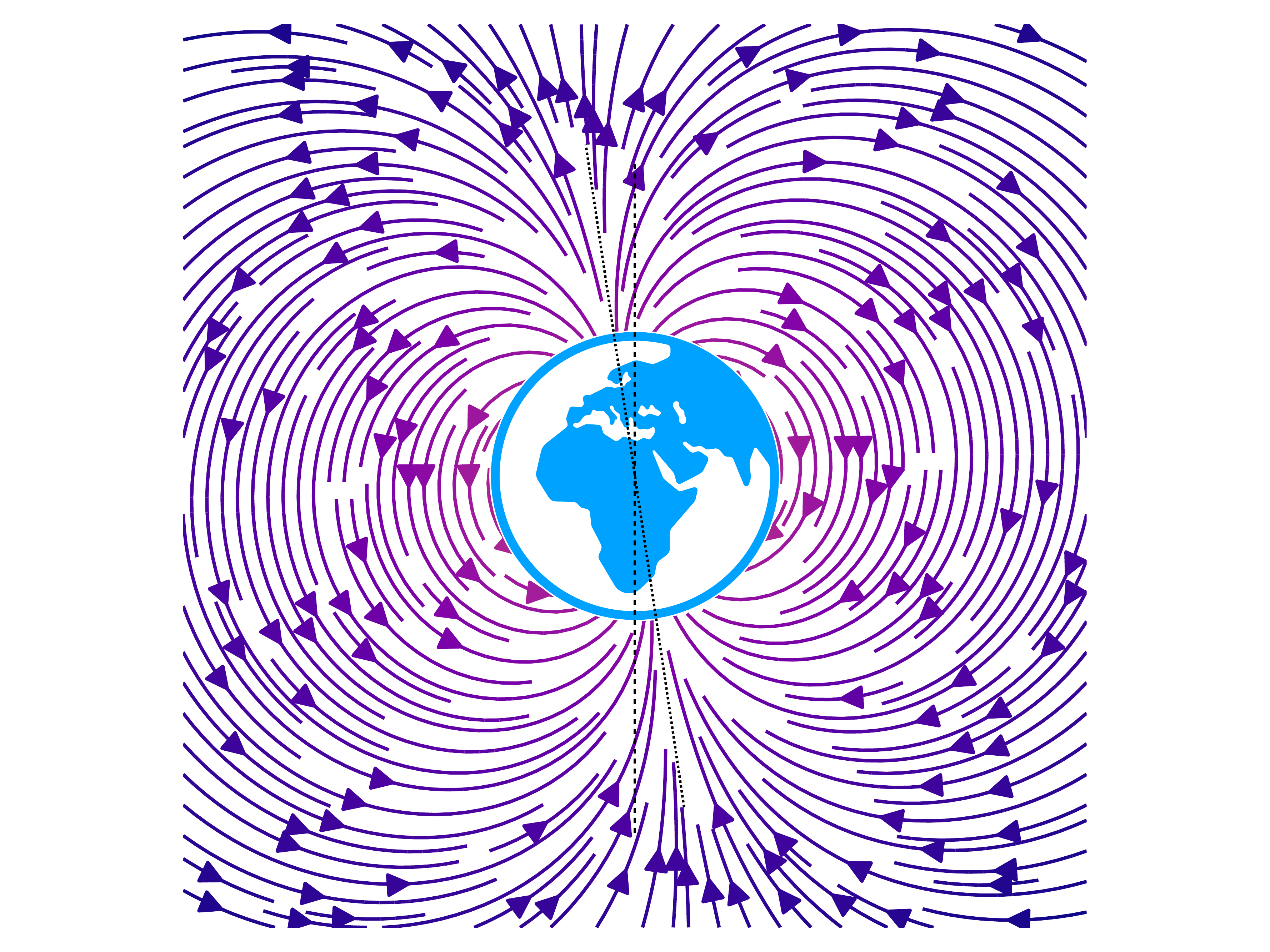}
\caption{The geomagnetic field, in the dipolar approximation.}
\label{f:dipole}
\end{figure}

The Earth is surrounded by the geomagnetic field, produced by the convection currents in the molten outer core. At Earth surface, its magnitude is about 0.5~G (50,000~nT).
As a first approximation (good especially at lower altitudes) the geomagnetic field can be described by an inclined dipole (Figure~\ref{f:dipole}), offset from Earth center by about 500~km, and tilted by about 10$^\circ$ with respect to the rotation axis. 
More complex and accurate models, which include also the dynamical variation of the geomagnetic field characteristics, are given for example by an expansion in spherical harmonics (e.g., the International Geomagnetic Reference Field, IGRF \cite{alken21}). 

\begin{figure}[htbp]
\centering
\includegraphics[width=\textwidth]{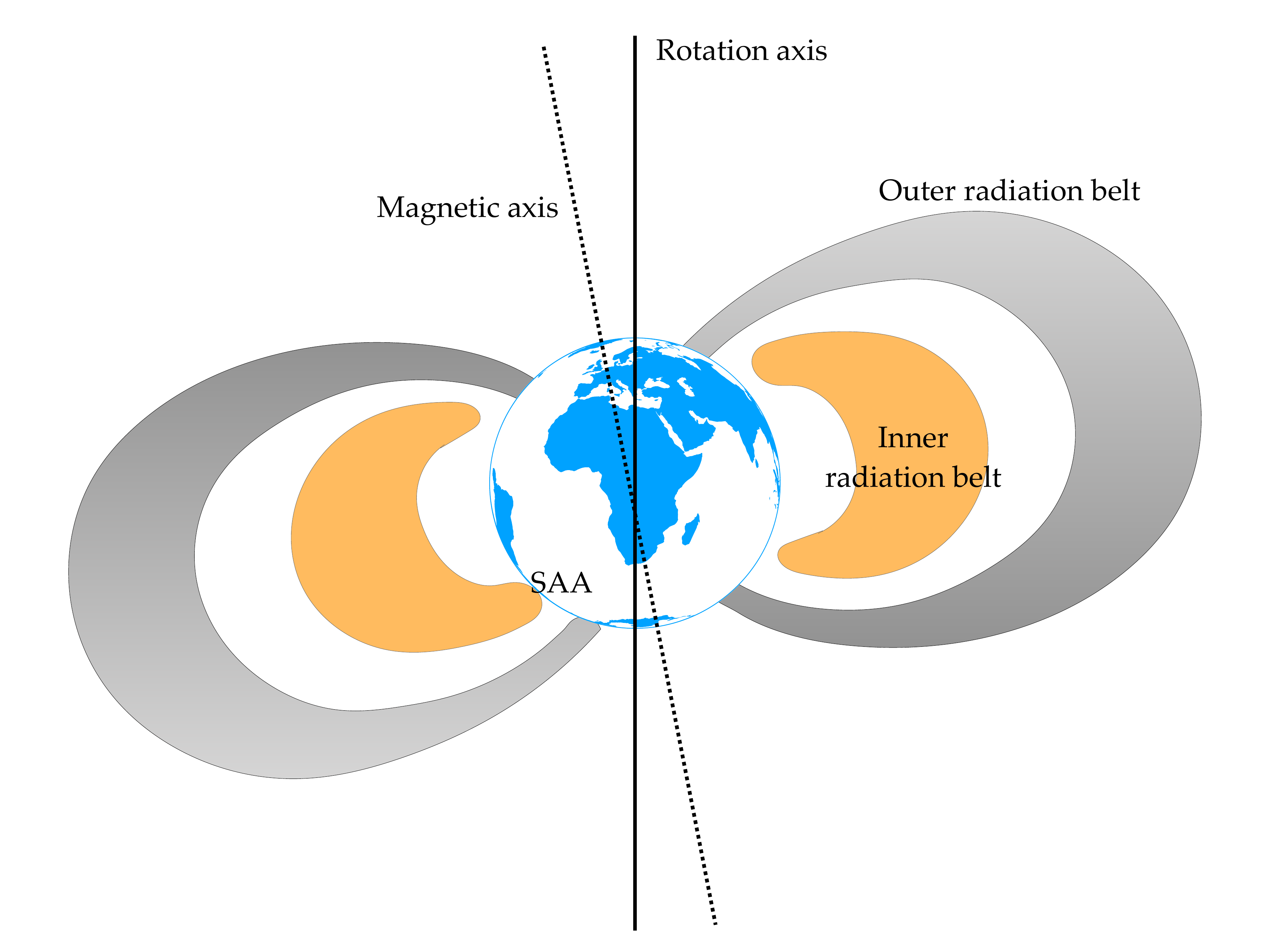}
\caption{The Van Allen radiation belts.}
\label{f:vanallen}
\end{figure}

At high altitudes, the geomagnetic field interacts with and is shaped by the solar wind. One result of the complex interplay between solar wind, cosmic rays and the geomagnetic field lines is the formation of two regions in which charged particles are trapped, the \emph{Van Allen radiation belts} (Figure~\ref{f:vanallen}).
Usually, an inner and an outer belt are defined. The inner belt extends from an altitude of about 1000 to 12,000 km, and contains mostly high-energy protons (MeV to GeV energies) and electrons up to a few MeVs. The outer belt extends from about 12,000 to 60,000 km and it is composed mostly of high energy electrons (hundreds of MeVs).

The South Atlantic Anomaly (SAA) is an area, approximately above the coasts of Brazil and of the southern Atlantic ocean, in which the Earth's inner Van Allen radiation belt comes closest to the Earth's surface, down to an altitude of 200 km.
The effect is caused by the non-concentricity of the Earth and its magnetic dipole, and the SAA is the near-Earth region where the magnetic field is weakest relative to an idealised Earth-centred dipole field.
This leads to an increased flux of energetic particles in this region and exposes orbiting satellites to higher-than-usual levels of radiation.

\subsection{Trapped particles}

The geomagnetically trapped electrons and protons inside the inner van Allen radiation belt are an important contributing factor for the overall instrumental background for a X-ray detector orbiting the Earth in a LEO or a HEO, i.e., in a path that crosses the SAA, the polar regions and/or the full radiation belt once per orbit.

While usually observations are not carried on during a SAA transit due to the huge rate of background events originating from the high flux of charged particles (e.g., for BeppoSAX/PDS \cite{frontera97} the high voltage of the instrument photomultiplier tubes was reduced, while other instruments could be completely switched off) an evaluation of the trapped particle flux is particularly important, given their contribution to the radio-activation of the instruments and spacecraft structures and to the radiation damage of the detector itself.  

Over the last decades several models describing the fluxes of the trapped particles have been developed, based on in-situ measurements by an array of different instrument and missions.

\begin{figure}[htbp]
\centering
\includegraphics[width=\textwidth]{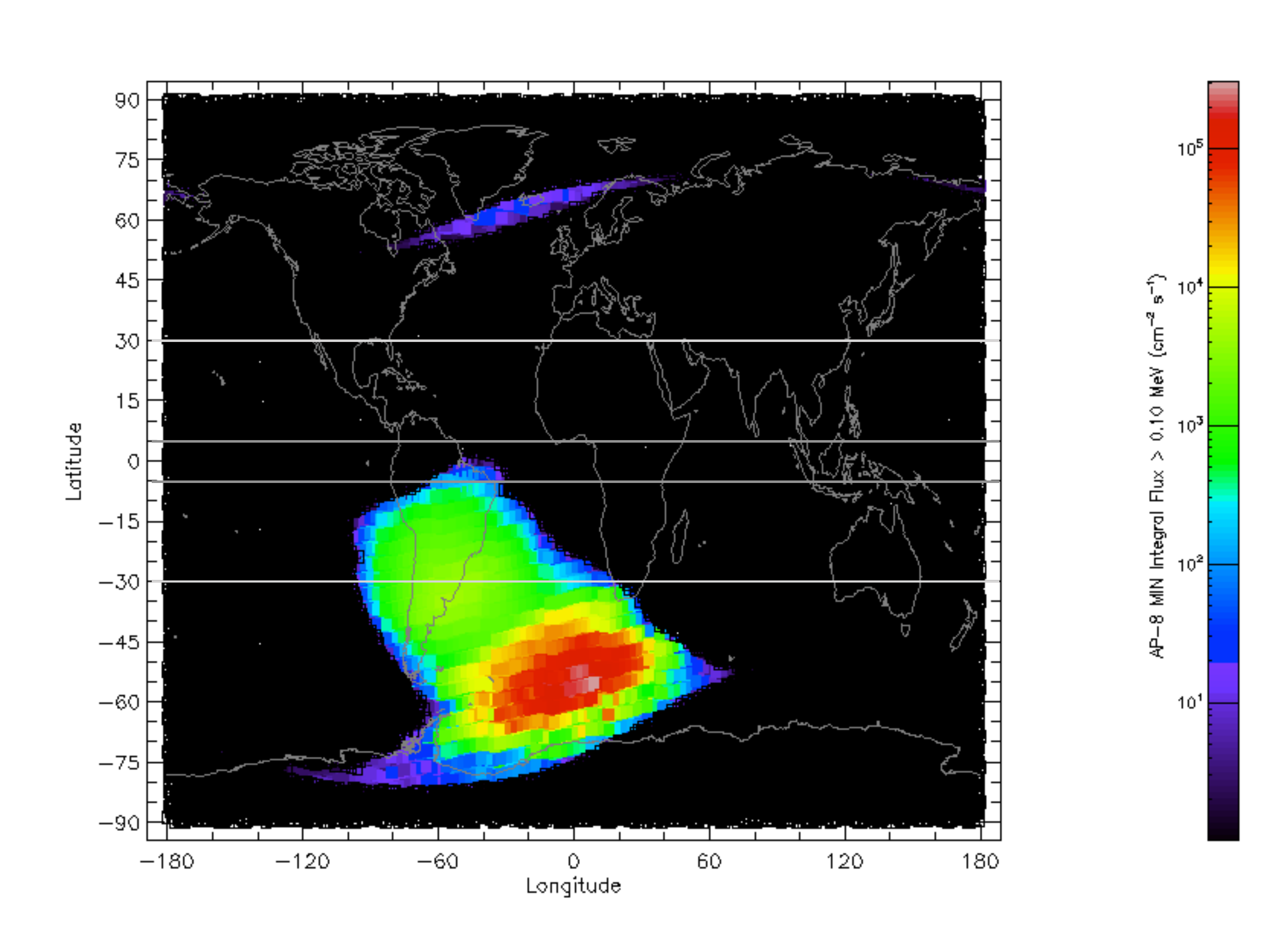}
\caption{Map of the AP-8 (solar minimum) trapped proton integral fluxes, above 100 keV, at an altitude of 600 km. The region with the strongest emission is the SAA. The gray lines mark geographic latitudes $\pm5^\circ$ and $\pm30^\circ$. Source: SPENVIS.}
\label{f:fluxp_map}
\end{figure}

The models which  have been usually adopted as a \emph{de-facto} standard in the aerospace industry are the NASA-developed AP-8 \cite{sawyer76} and AE-8 \cite{vette91} models, based on data from several satellites flown in the '60s--'70s. These models provide omnidirectional, integral electron (AE) and proton (AP) fluxes in the  range between 0.04 and 7~MeV for electrons, and 0.1--400~MeV for protons (Figure~\ref{f:fluxp_map}). Models are available for solar minimum and maximum conditions, but do not involve any modelling of the variable geomagnetic field conditions. Moreover, large uncertainties are present for the predicted fluxes in the regions where steep gradients in spatial and spectral distribution exist (e.g., the boundaries of the SAA at low latitudes and altitudes).

In the last thirty years several new models have been developed. For example, a more updated low altitude ($<$600 km) trapped proton model for solar minimum conditions has been developed in the framework of an ESA/ESTEC sponsored project,  based on measurements made by the Proton/Electron Telescope (PET) onboard the SAMPEX satellite \cite{heynderickx99}.

Recently, a dynamic update of the AP8/AE8 models has been developed by the National Reconnaissance Office (NRO) and the Air Force Research Laboratory (AFRL), resulting in the AE9/AP9/SPM models \cite{ginet13} based on more than 30 satellite data sets spanning the years from 1976 to 2011. These models include also space plasmas, and offer significant improvements over AP8/AE8, such a higher spatial resolution and the quantification of uncertainties due to both space weather and instrument errors, by using among others a Monte-Carlo approach to compute flux thresholds for arbitrary percentile levels (e.g., 50th and 95th). Several inconsistencies and uncertainties remain, however, in particular for low-inclination and low-altitude environments.

The \emph{SPace ENVironment Information System} (SPENVIS) platform\footnote{\url{https://www.spenvis.oma.be}, developed by a consortium led by the
Royal Belgian Institute for Space Aeronomy (BIRA-IASB).} is an Internet-based interface to various models of the space environment and its effects on detectors and electronics, including cosmic rays, natural radiation belts, solar energetic particles, plasmas, gases, and so on.
A computation of the expected trapped radiation environment for a given satellite orbit can be then performed using the above mentioned models.

As shown in Figure~\ref{f:fluxp_map}, a satellite in a low inclination orbit (e.g., with an orbital inclination of 5$^\circ$) will graze the SAA only in its outermost regions, with typical smaller transit durations and much lower trapped particle fluxes with respect to a satellite in a higher inclination (e.g., 30$^\circ$) which would cross the ``core'' of the SAA.

\subsection{Solar particles}

Besides the solar wind, i.e., the continuous stream of soft charged particles (electrons, protons, ions with energies typically below 10 keV) produced and accelerated in the solar corona by heating and magnetic fields, the Sun emits also the so-called \emph{solar energetic particles} (SEP), for which the kinetic energies can range from 10 keV to several GeVs. These are produced by solar flares and in shock regions associated with coronal mass ejections. As such, these bursts of high energy accelerated particles have strongly variable fluxes and fluences, extremely difficult to predict and that can be very challenging for space missions, especially for their sensitive electronics. The geomagnetic field provides some shielding, in particular at low altitudes and inclinations, so SEP events are a concern mostly for missions in HEOs or at $L_2$ (especially when carrying X-ray optics which can focus also soft protons, see Section~\ref{s:charged}).

There are several models for estimating SEP fluences and maximum fluxes, useful in order to optimize the design of a detector and its electronics.
However, given the random nature of these events, the usual approach is to filter out from the data the time intervals in which solar flares occur.

\subsection{Cosmic rays}

The cosmic rays, mostly composed by high-energy protons ($\sim$85\%), ions ($\sim$13\%) and electrons ($\sim$2\%) \cite{longair06}, are particles accelerated to relativistic velocities in different processes, originating from Galactic and extra-Galactic sources. Interacting with the atmosphere, cosmic rays produce showers of secondary particles, which can reach the Earth's surface but can be also reflected back into space.

The \emph{primary} cosmic ray spectrum extends from an energy of a few MeV/nucleon to extremely high energies, above 10$^{20}$ eV/nucleon. The differential spectrum is well described, above $\sim$1~GeV/nucleon, by a power-law distribution, $F(E) = dN/dE \propto E^{-\gamma}$, where $\gamma \sim 2.7$.
The intensity of the cosmic ray flux  is modulated by the 11-year solar cycle, with a variation in the total integral flux of cosmic rays at the Earth location of about a factor of two between solar minimum and maximum. The magnetic fields transported by the solar wind can deflect and scatter the incoming cosmic rays. One parameter that quantifies the effect of the solar cycle is the \emph{solar modulation potential} $\Phi$ \cite{gleeson68}, usually measured in GV.

The presence of the geomagnetic field, moreover, modifies the trajectories of the charged particles near the Earth. The relevant quantity here is the particle \emph{rigidity} $R = pc/q$, where $p$ is the  momentum, $q$ its charge and $c$ is the speed of light, i.e., the momentum for unit charge (measured in volts).  If an individual cosmic ray has a rigidity above a certain threshold, it can reach a given location in the magnetosphere. This threshold value is called \emph{cutoff rigidity}, $R_\mathrm{cut}$, and is a complex function of the altitude, of the geomagnetic latitude and of the incoming direction of the incident particle. A simple approximation is the Størmer formula \cite{smart05}, which gives the cutoff rigidity for vertically incident particles: 
\begin{equation}
R_\mathrm{cut} = 14.5 \times \left( 1 + \frac{h}{R_E} \right)^{-2} \cos^4 \lambda \mbox{\,\,\, GV}
\end{equation}
where $\lambda$ is the geomagnetic latitude, $h$ the altitude and $R_E = 6371$~km is the Earth radius. For LEO altitudes, the vertical cutoff rigidity ranges from $\sim$12 GV at the geomagnetic equator to $\leq$3 GV at mid-latitudes.

By taking into account all these factors, the full differential spectrum for a primary cosmic ray particle of mass $M$, charge $Ze$, rigidity $R$ and kinetic energy $E$, at a location around the Earth given by an altitude $h$ and a geomagnetic latitude $\lambda$, can be expressed by:
\begin{equation}
F(E) = F_U(E) \times F_M(E, M, Z, \Phi)  \times C(R, h, \lambda)
\end{equation}
in which $F_U$ is the incoming, unmodulated, cosmic ray spectrum, $F_M$ is an analytic solar-cycle modulation factor and $C$ is a geomagnetic cutoff function \cite{mizuno04,campana13}.

The unmodulated cosmic ray flux can be described by several models. A simple approximation \cite{mizuno04}, based on AMS measurements \cite{alcaraz00a, alcaraz00b}, is
\begin{equation}
F_U(E) = K \times \left[ \frac{R(E)}{\mathrm{GV}} \right]^{-\gamma}
\end{equation}
where $R(E)$ is the rigidity\footnote{Rigidity can be expressed as a function of the kinetic energy $E$ per nucleon as $R(E) = \frac{A}{Z} \sqrt{E^2 + 2Mc^2E}$, where $A$ is the number of nucleons, $Z$ the charge and $M$ the mass.} in GV
and $K$ and $\gamma$ are parameters which depend on particle type and geomagnetic latitude. For example, for protons $\gamma = 2.83$ and $K = 23.9$ particles m$^{-2}$ sr$^{-1}$ MeV$^{-1}$. 

The geomagnetic cutoff function can be expressed with the simplified relation
\begin{equation}
C(R, h, \lambda) = \frac{1}{1 + \left(\frac{R}{R_\mathrm{cut}}\right)^{-r}}
\end{equation}
where $R_\mathrm{cut}$ is the vertical Størmer cutoff rigidity and the parameter $r$ is 12 for protons and 6 for leptons.

The SPENVIS platform allows to compute the cosmic-ray spectrum for a given orbit with more complex models, such as the ISO-15390 standard and the CREME96 model \cite{tylka97}, taking into account also the short-time variations of the magnetosphere (geomagnetic storms) and employ a more accurate numerical calculation of the cutoff rigidity.
Usually, results are compatible within a factor of few \cite{cumani19, galgoczi21} with respect to the simplified approach presented here.

The interaction of cosmic rays with the Earth atmosphere produces secondary particles, which can also be ``quasi-trapped'' in the magnetic field lines. Observation performed by AMS \cite{alcaraz00a,alcaraz00b} and PAMELA \cite{adriani15} showed that there are different populations of particles with different lifetimes. Simplified analytic models (with a power-law, broken power-law or cutoff power-law shape) for different geomagnetic latitudes, different energies and for upward and downward fluxes are available \cite{mizuno04, campana13, galgoczi21}, based on AMS \cite{alcaraz00a,alcaraz00b}, MITA/NINA-II \cite{bidoli02} and MIR/Maria-2 \cite{mikhailov02} measurements.

\subsection{Neutron albedo radiation}

Hadronic showers induced by the interaction of primary cosmic rays with the atmosphere can also produce neutrons, which are able to reach LEO altitudes before decaying. The resulting neutron flux will depend on the geomagnetic latitude, since it origins from the downward-incident primary cosmic-ray population, with a flux increase of about an order of magnitude for polar regions with respect to equatorial latitudes. Moreover, scattering effects with atmospheric nuclei will slow down (``thermalise'') a significant fraction of neutrons, giving rise to an output spectrum ranging from eV energies to several GeVs.

Currently, there are several models describing the neutron albedo radiation for a given altitude and geomagnetic latitude, mostly based on Monte Carlo transport codes \cite{armstrong73, kole15}, and validated with balloon-borne and satellite-based (e.g., Comptel/CGRO \cite{morris98}) observations.

The latest revision of the ESA standard ECSS-E-ST-10-04C \cite{ECSS-E-ST-10-04C} suggests the use of the MAIRE (Models for Atmospheric Ionising Radiation Effects\footnote{\url{http://www.radmod.co.uk/maire}}) code, which is an update of the QinetiQ Atmospheric Radiation Model (QARM) that has been extensively validated \cite{lei04,lei06}.

\subsection{Cosmic X-ray diffuse background}

The principal source of photon background for a space-borne X-ray astronomy mission is the nearly isotropic diffuse cosmic radiation (cosmic X-ray background, CXB), that is, the integrated emission of extragalactic point sources such as active Galactic nuclei (AGN) powered by supermassive black holes. Discovered during the very first X-ray astronomy experiments in the '60s \cite{giacconi62}, the CXB is dominated by the emission from quasars and Seyfert galaxies at low energies, while above $\sim$10 keV the main contribution is from highly absorbed AGNs. An imaging instrument with sufficient angular resolution and sensitivity can resolve a substantial fraction of the CXB into individual sources, while for collimated or wide-field instruments a total emission can be considered. The Poisson statistics of point-like field sources lead to the \emph{cosmic variance}, i.e., a statistical fluctuation of the flux integrated on a given solid angle.

The ``canonical'' model for the CXB emission spectrum above $\sim$2 keV is given by Gruber et al. (1999) \cite{gruber99}, based on HEAO-1/A4 and COMPTEL measurements, and substantially confirmed by other observations \cite{frontera07, churazov07, ajello08, krivonos21}.

The CXB spectrum, peaking at around 30 keV, is described in this model by different functions in different energy ranges. The first branch, valid below 60 keV, is:
\begin{equation}
F(E) = 7.877 \times \left( \frac{E}{1 \mathrm{\, keV}} \right)^{-1.29} e^{-E/41.13} \mbox{\, photons cm$^{-2}$ s$^{-1}$ sr$^{-1}$ keV$^{-1}$}
\end{equation}
while for energies above 60 keV:
\begin{eqnarray}
F(E) &= 0.0259  \times \left( \frac{E}{60 \mathrm{\, keV}} \right)^{-5.5} \nonumber\\& +  0.504\times \left( \frac{E}{60 \mathrm{\, keV}} \right)^{-1.58}  \\& +   0.0288  \times \left( \frac{E}{60 \mathrm{\, keV}} \right)^{-1.05}   \nonumber\\   & \mbox{\, photons cm$^{-2}$ s$^{-1}$ sr$^{-1}$ keV$^{-1}$}\nonumber
\end{eqnarray}

Measuring the absolute intensity of the CXB is a thorny problem, linked to the absolute calibration and cross-calibration of a X-ray detector and to its background rejection efficiency.
By comparing several measurements from different experiments, the uncertainty on the absolute intensity is about 20\%. However, there is some indication \cite{ajello08, krivonos21} that the absolute flux at the CXB $\sim$30 keV peak should be somewhat ($\sim$8\%) higher than the one described by the above mentioned model.

\subsection{Galactic diffuse emission}
Besides the extragalactic component (CXB), the diffuse photon emission includes also a Galactic contribution, non-spatially uniform and due to the integrated contribution of different classes of sources (e.g., cataclysmic variables, coronally active binaries, \cite{revnivtsev06,turler10}).  The emission is strongest in the central radians of the Galaxy (i.e., approximately for absolute Galactic longitudes below 30$^\circ$ and latitudes below 10$^\circ$), but a thin component all along the Galactic disk is still present at other longitudes.

\subsection{Earth gamma-ray albedo radiation}

The interaction of the primary cosmic-rays with the Earth atmosphere produces, among a shower of secondary particles, also a X- and $\gamma$-ray photon population, mainly produced through bremsstrahlung and $\pi^0$ decay processes. Moreover, the diffuse X-ray emission can be reflected by the atmosphere itself, dominating the total atmospheric emission below a few tens of keV \cite{churazov08} .

This Earth albedo emission intensity is strongly dependent on the primary flux modulated by the local geomagnetic conditions, and therefore on the latitude. 

Several data-based and simulation-based models have been developed \cite{sazonov07,churazov08} to describe the Earth gamma-ray albedo radiation inside and outside the atmosphere.  For example, for a low altitude/low inclination environment a model based on \emph{Swift}/BAT measurement has been developed by Ajello et al. (2008) \cite{ajello08}, consistent with previous observations, that can be parameterised as:
\begin{equation}\label{e:albedo}
F(E) =   \frac{0.0148}{\left(\frac{E}{33.7 \mathrm{\, keV}}\right)^{-5} + \left(\frac{E}{33.7 \mathrm{\, keV}}\right)^{1.72}} \mbox{\, photons cm$^{-2}$ s$^{-1}$ sr$^{-1}$ keV$^{-1}$}
\end{equation}

\medskip

\begin{figure}[htbp]
\centering
\includegraphics[width=\textwidth]{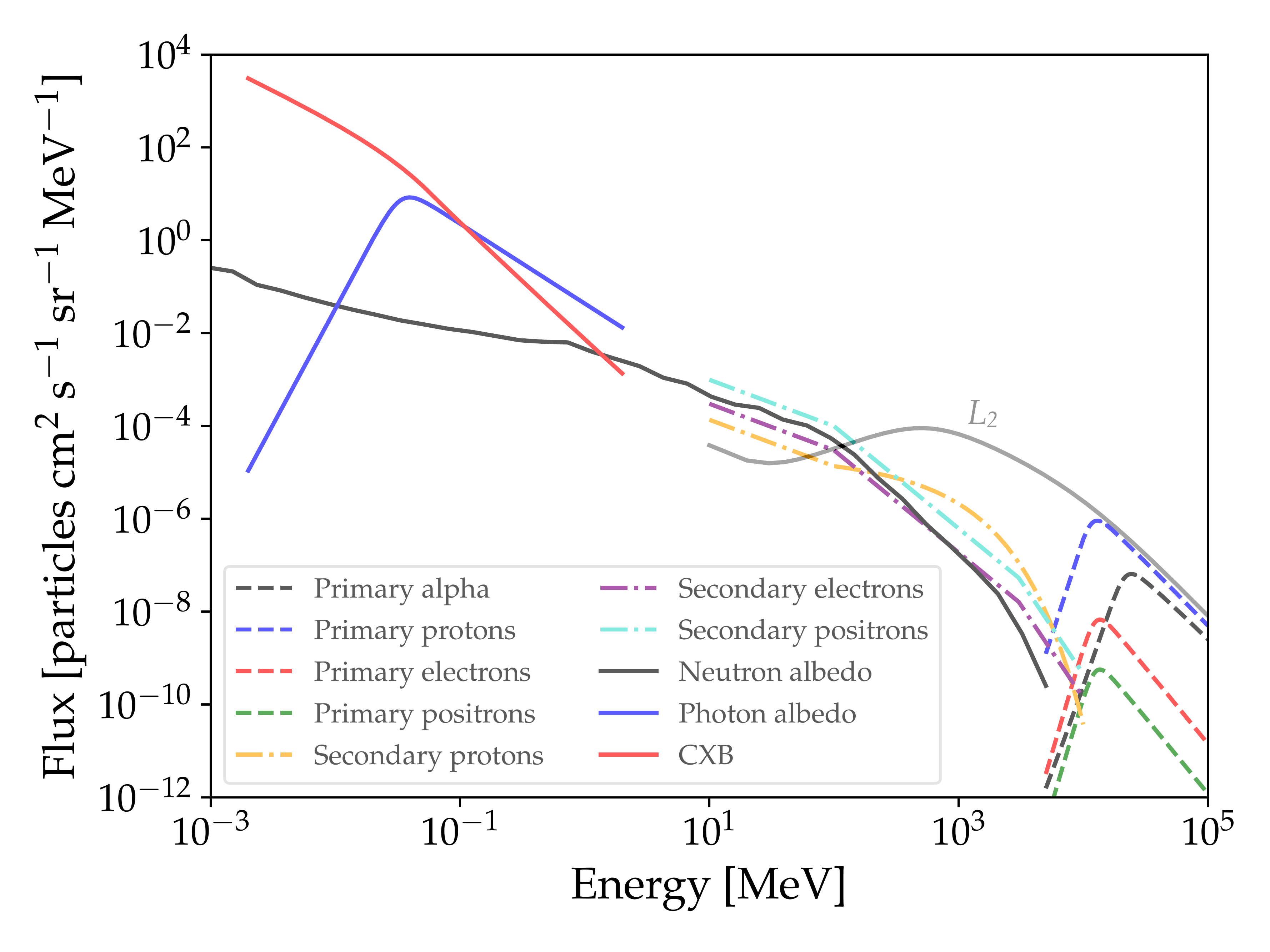}
\caption{The various background sources in an equatorial LEO. For comparison, the primary cosmic-ray proton spectrum for a $L_2$ orbit is also shown.}
\label{f:bkg_src}
\end{figure}

Figure~\ref{f:bkg_src} shows a summary of the spectra of all the above mentioned background contributions expected for an equatorial LEO (i.e., with an orbital inclination below $\sim$5$^\circ$) at altitudes typical of a X-ray observatory (500--600 km). Note the effect of the geomagnetic cutoff around GeV energies for the primary cosmic ray populations, and the fact that in X-rays the Earth becomes brighter than the rest of the sky, per unit of solid angle, above around 100 keV. 

Given that a spacecraft orbits around the Earth likely spanning different local geomagnetic conditions (e.g., cutoff rigidities) and often with a continuously changing attitude with respect to the Earth, the interplay of these factors make the instantaneous local background environment a dynamic quantity. However, average spectra and fluxes as discussed above and shown in Figure~\ref{f:bkg_src} are useful as a starting point to estimate the expected background level when designing an instrument.

\section{Radiation effects on detectors}\label{s:radeffects}
\subsection{Radiation damage}

The space environment can have different effects on a satellite-borne X-ray instrument: first and foremost, the interaction of high-energy particles can damage the detector itself. One mechanism relevant for solid-state detectors, for example, is the so-called \emph{non-ionizing energy loss} (NIEL), in which the transfer of energy from an incoming particle can damage the crystalline lattice of a semiconductor detector: such a damage will cumulate with time, leading to an increasing \emph{dark current} which has impacts on the detector performance.
Another typical process, important in particular for the CCDs, is the \emph{charge transfer efficiency} (CTE) decrease, i.e., the efficiency in transferring a photoelectron, from one pixel to the next, when reading out the device. Low CTE values can decrease the energy resolution, and moreover ``hot'' pixels or ``streaks''  spanning one column of the CCD matrix can be induced.

The design of an experiment shall take these effects into account, leading to constraints on detector requirements (e.g., operating temperature) and expected lifetime (i.e., the maximum time in which the detector is able to satisfy image quality or spectral resolution requirements).

\subsection{Scientific background effects}

In the following we will focus on the immediate effects of the interaction of the background sources with space-borne X-ray experiments most relevant for the analysis of the scientific data from a high-energy astrophysics mission.

In general, we can identify three main areas in which the environment can contribute to the instrumental background:
\begin{itemize}
\item \emph{Timing}. Variable background sources can impact the detected count rate as a function of time, with both short-term (e.g., flares) and long-term effects (e.g., average count rate level drifts over minutes or hours, correlated with local orbital phase and/or satellite attitude)

\item \emph{Spectral}. Background sources can leave their imprint in the detected spectra, both as an additional continuum or line contribution (e.g., by inducing fluorescence in the detector materials themselves)

\item \emph{Imaging}. For an imaging instrument, background sources can contribute to local effects (e.g., ``hot'' pixels or ionisation streaks) and detector non-uniformities.

\end{itemize}

All these effects, ultimately, derive from the details of the interactions between a given background particle with the detector and its surrounding structure. Different sources (e.g., photons or particles) will have different effects, and a correct understanding of their impact on the scientific observations will depend on the detail of each experiment.

\subsection{Photon background}

Focusing and collimated instruments, aimed to observe a point-like cosmic source, will collect also the diffuse cosmic emission, whose photons will sum to those originating from the target and in principle cannot be recognised as such (a photon is a photon, regardless of its origin!). In order to correctly evaluate the photometry and the spectrum of the source of interest, the background contribution has to be evaluated and subtracted.

\begin{figure}[htbp]
\centering
\includegraphics[width=\textwidth]{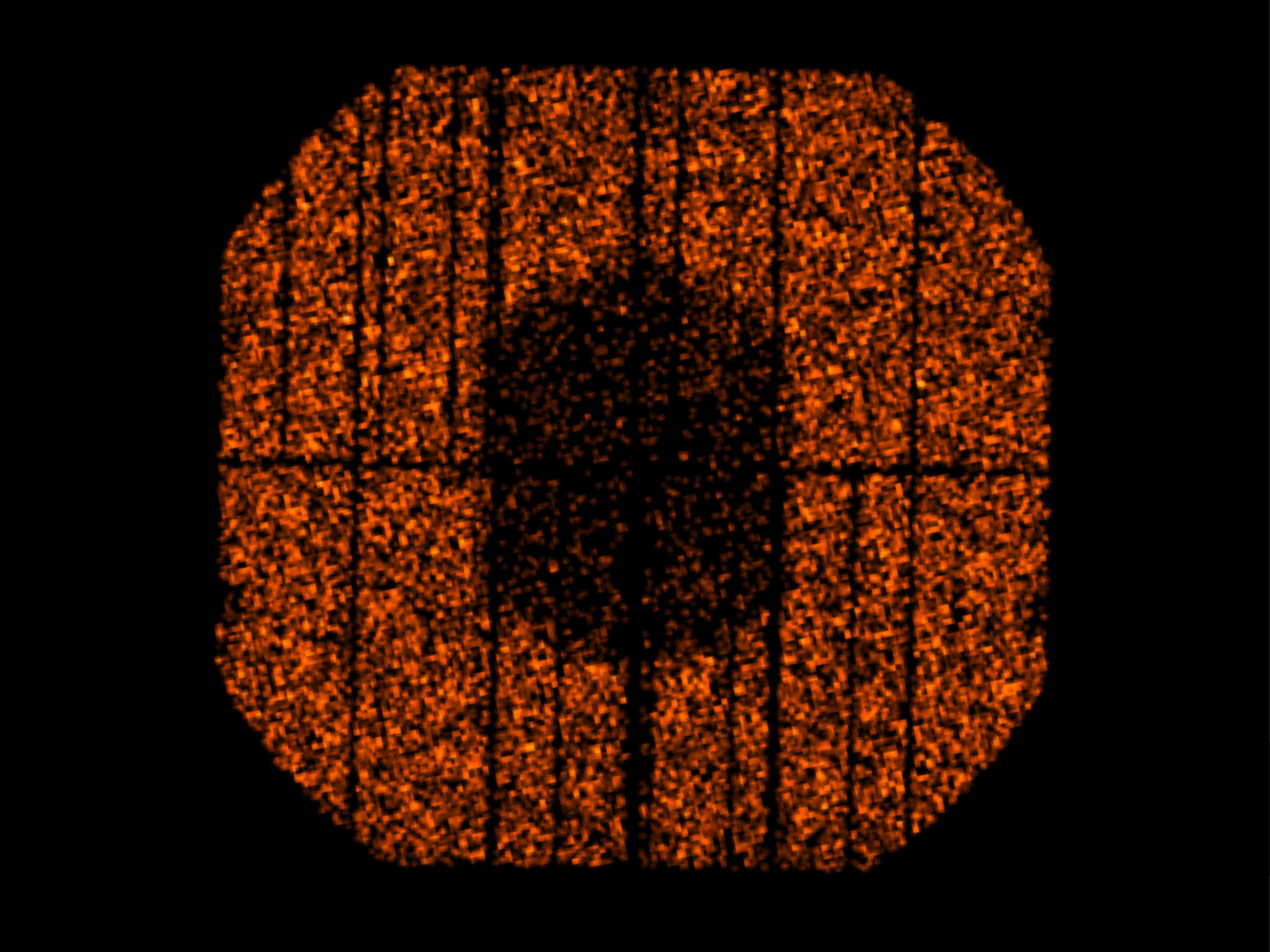}
\caption{One XMM-Newton EPIC-pn image, selected in the 7.8--8.2 keV range, showing the effect of the fluorescence of Cu layers behind the detector.}
\label{f:epicpn_fluo}
\end{figure}

Besides contributing to the overall instrumental noise, the diffuse X-ray photons (and also particles) can induce \emph{fluorescence} in the detector structures. One example is reported in Figure~\ref{f:epicpn_fluo}, which shows the spatially non-homogeneous fluorescence induced by high energy photons or cosmic rays in the copper layers behind the XMM-Newton EPIC-pn detector ($K_\alpha$ line at 8.05 keV).

\subsection{Charged particles}\label{s:charged}
Charged particles (hadrons or leptons) can interact with the satellite and the instrument in complex ways. Usually, they leave in their wake an \emph{ionisation streak}, i.e., a continuous ionisation of the material along their path. These electrical charges can be detected by the instrument. 

The mean energy loss per unit path length by a charged particle of charge $ze$, energy $E$ and speed $v$ is described by the Bethe-Bloch formula:
\begin{equation}
\frac{dE}{dx} = \frac{z^2e^4N_e}{4\pi\epsilon_0^2m_ev^2} \left[ \log \left( \frac{2\gamma m_e v^2}{I} \right) - \frac {v^2}{c^2} \right]
\end{equation}

Typically, a particle in the space environment can be treated as a \emph{minimum ionizing particle} (MIP), i.e., the ionisation along its path is comparable to the minimum value given by the equation above.
The amount of ionisation, and thus the energy deposit by a charged particle on a detector, will depend on the material. As a rule of thumb, for a MIP crossing a silicon layer the mean energy loss is $dE/dx = 3.87$ MeV/cm. This means that when crossing typical solid-state detector thicknesses (300--500 $\mu$m) a particle will deposit around 100--200 keV. For much thinner detectors (such as the transition edge sensor, TES, micro-calorimeters) the energy deposit for a MIP could fall within the nominal sensitivity range.

\begin{figure}[htbp]
\centering
\includegraphics[width=\textwidth]{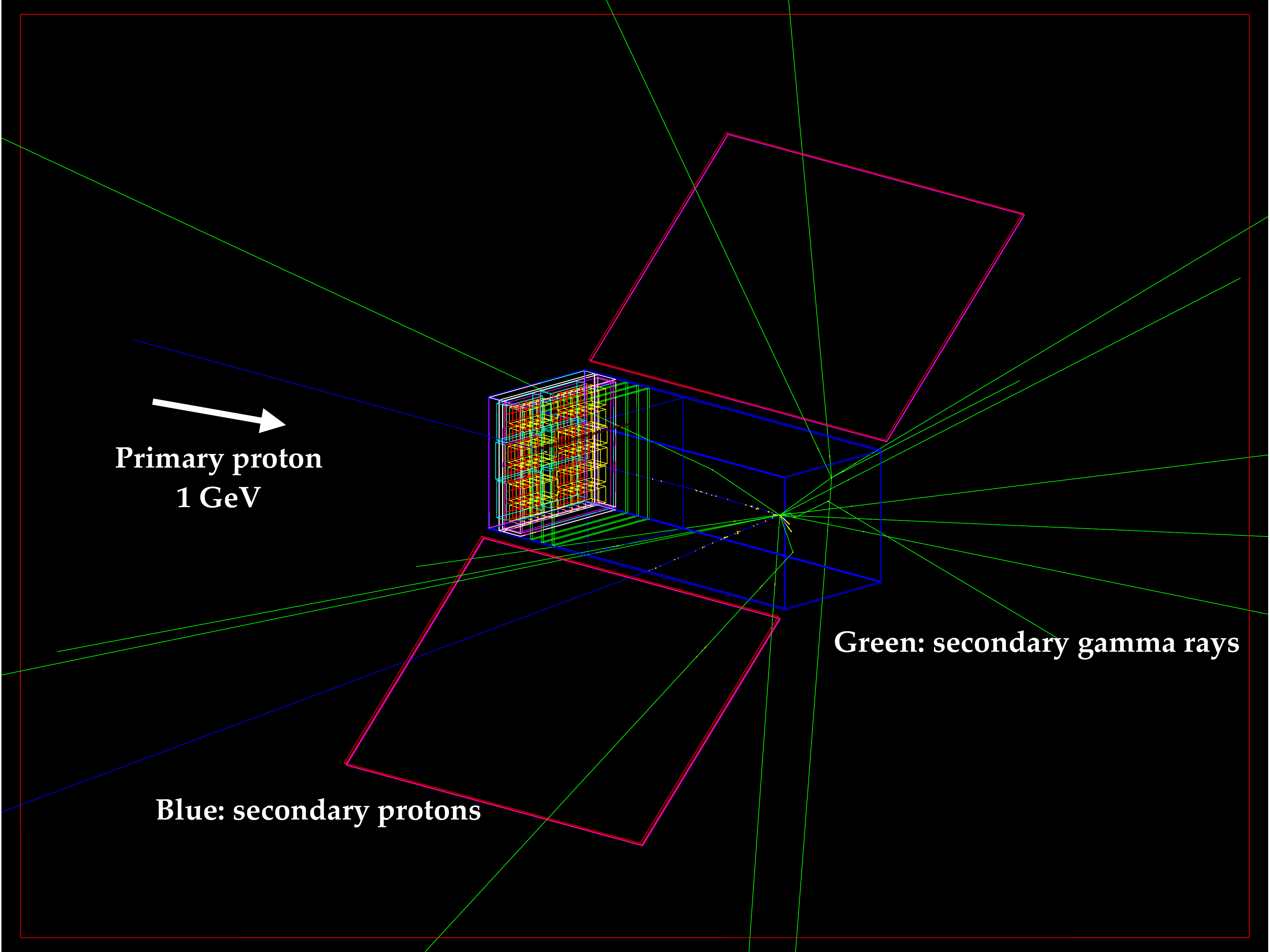}
\caption{Monte Carlo simulation of the interaction of a 1 GeV primary proton incident on a CubeSat satellite carrying an high-energy detector.}
\label{f:exampleMC}
\end{figure}

Another process which produces charged particles is the production of secondary showers by the electromagnetic and hadronic interactions of an high-energy primary with the instrument or the spacecraft structures.  
Figure \ref{f:exampleMC} shows, as an example, a Monte Carlo simulation of a 1 GeV proton incident on a CubeSat satellite carrying an high-energy detector. The  interaction of the primary proton with the spacecraft produces several tracks of secondary protons and $\gamma$-rays, which can then escape the sensitive volume or interact further in complex ways with the detector.

\begin{figure}[htbp]
\centering
\includegraphics[width=\textwidth]{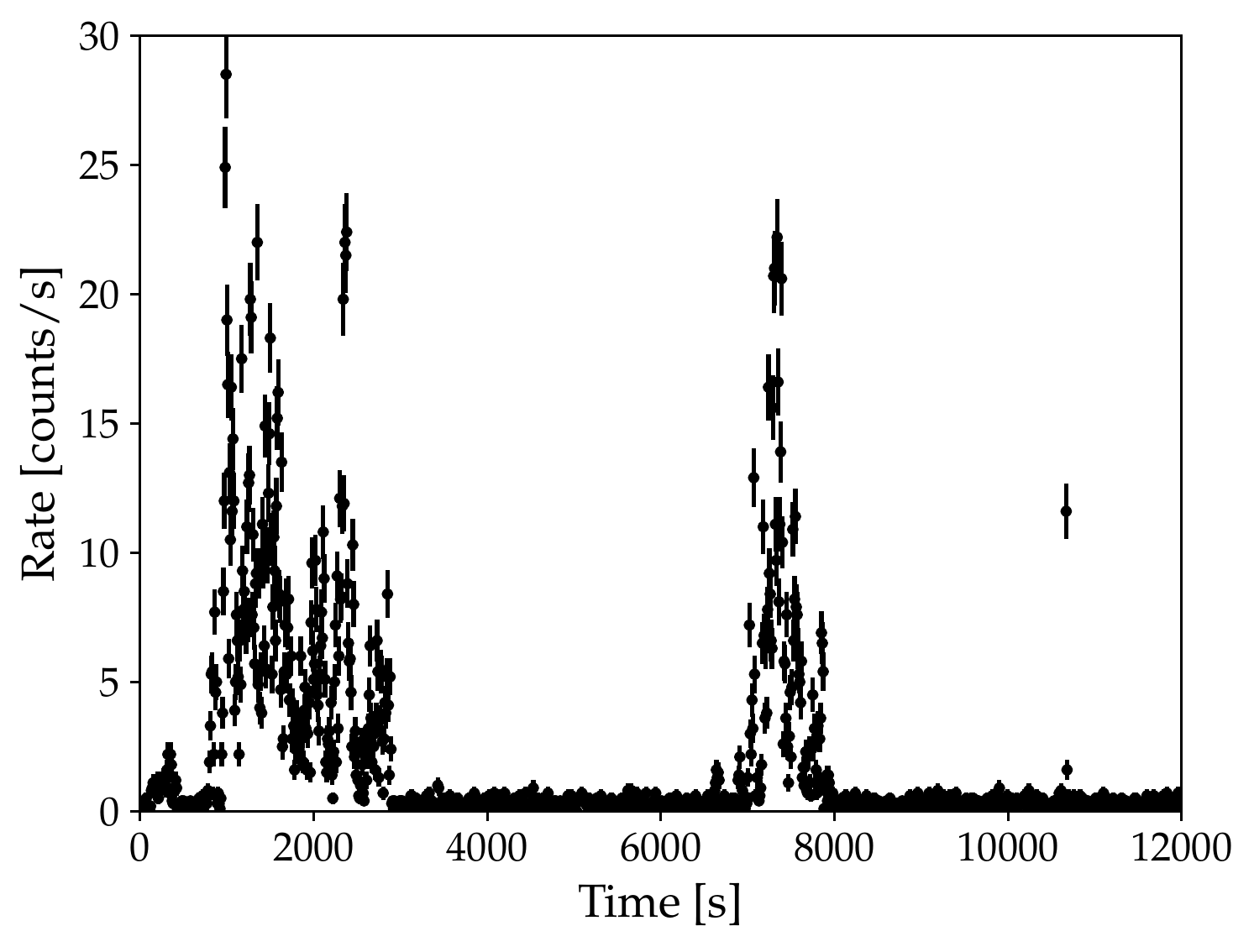}
\caption{One XMM-Newton EPIC-pn observation showing the effect of soft proton flares. The first $\sim$3000~s and a time interval around 7500~s from the start of the observation are badly affected by flares. These time intervals will be removed from the scientific analysis.}
\label{f:epicpn_lc}
\end{figure}

Low-energy protons, i.e., with energies below approximately 300~keV, of solar origin are present in the geomagnetically unshielded HEO/Lagrangian environment. These particles can be focused by X-ray optics, by grazing-incidence scattering processes similar to the photon ones \cite{fioretti17} resulting in \emph{flares} that can severely limit the useful observation time and contributing also to the detector damage. These soft proton flares can occur at random times, arising from complex phenomena in the dynamic Earth magnetosphere. As an example, for XMM-Newton (Figure~\ref{f:epicpn_lc}) about 40\% of the observation time is affected by this phenomenon, in a hard-to-predict manner.

\subsection{Activation}

High-energy protons encountered by the spacecraft along its orbit (for example, when crossing the trapped-particle regions, or by the bombardment of primary cosmic-rays) can induce \emph{radio-activation} of the various spacecraft and detector materials. Typical physical hadronic processes involve inelastic scattering, neutron capture and nuclear fragmentation, and the net result is the creation of various types of unstable isotopes, which can then decay with largely different processes and associated lifetimes, but ultimately performing as sources of unwanted $\gamma$-rays and ionizing particles.

\begin{figure}[htbp]
\centering
\includegraphics[width=\textwidth]{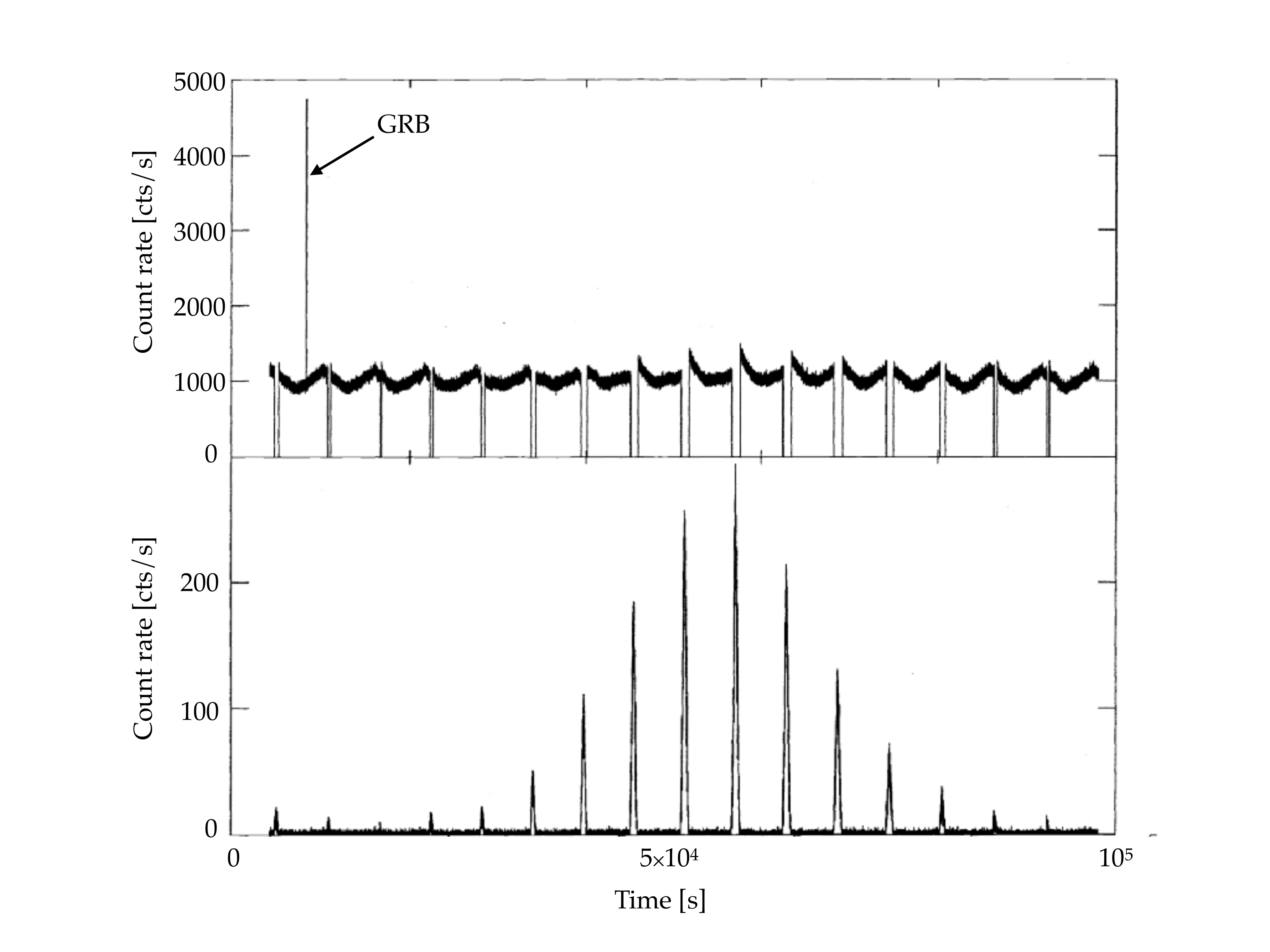}
\caption{Example of activation-induced background increase. Upper panel: count rate from the BeppoSAX/PDS anticoincidence (AC), which was composed by four CsI(Tl) scintillator slabs read out by a photomultiplier. Lower panel: Particle Monitor (PM) count rate. Data gaps in the AC rate correspond to the SAA transits, with various depths (peaks in the PM count rate). The correlation between transit depth and thus amount of high-energy proton irradiation, and short-term background increase due to the activation of the detector, is well apparent. The very short spike in the AC rate is due to a gamma-ray burst. Adapted from \cite{feroci97}.}
\label{f:feroci97}
\end{figure}

Therefore, the decay of the activation products will induce a strongly variable contribution to the background. The amount of the produced activation-induced radionuclides, with their half-lives and subsequent radioactive decay chains, will depend on the precise time profile of the irradiation (flux and spectrum). For example, a satellite orbiting in a LEO or in a HEO will be subject to both a short time-scale activation-induced background increase, due to the decay of nuclei with short lifetimes, just after the transit through trapped particle regions (see Figure~\ref{f:feroci97} for an example), and to a long (months/years) time-scale background increase, due to the accumulation of long-lived radioactive isotopes.

As such, the modelling of the activation is a rather complex task. Several approaches are possible, such as direct Monte Carlo simulations (e.g. using the MEGAlib toolkit, \cite{zoglauer06}) or the use of different, faster algorithms exploiting semi-analytical calculations of decay schemas \cite{odaka18}. 

\begin{figure}[htbp]
\centering
\includegraphics[width=\textwidth]{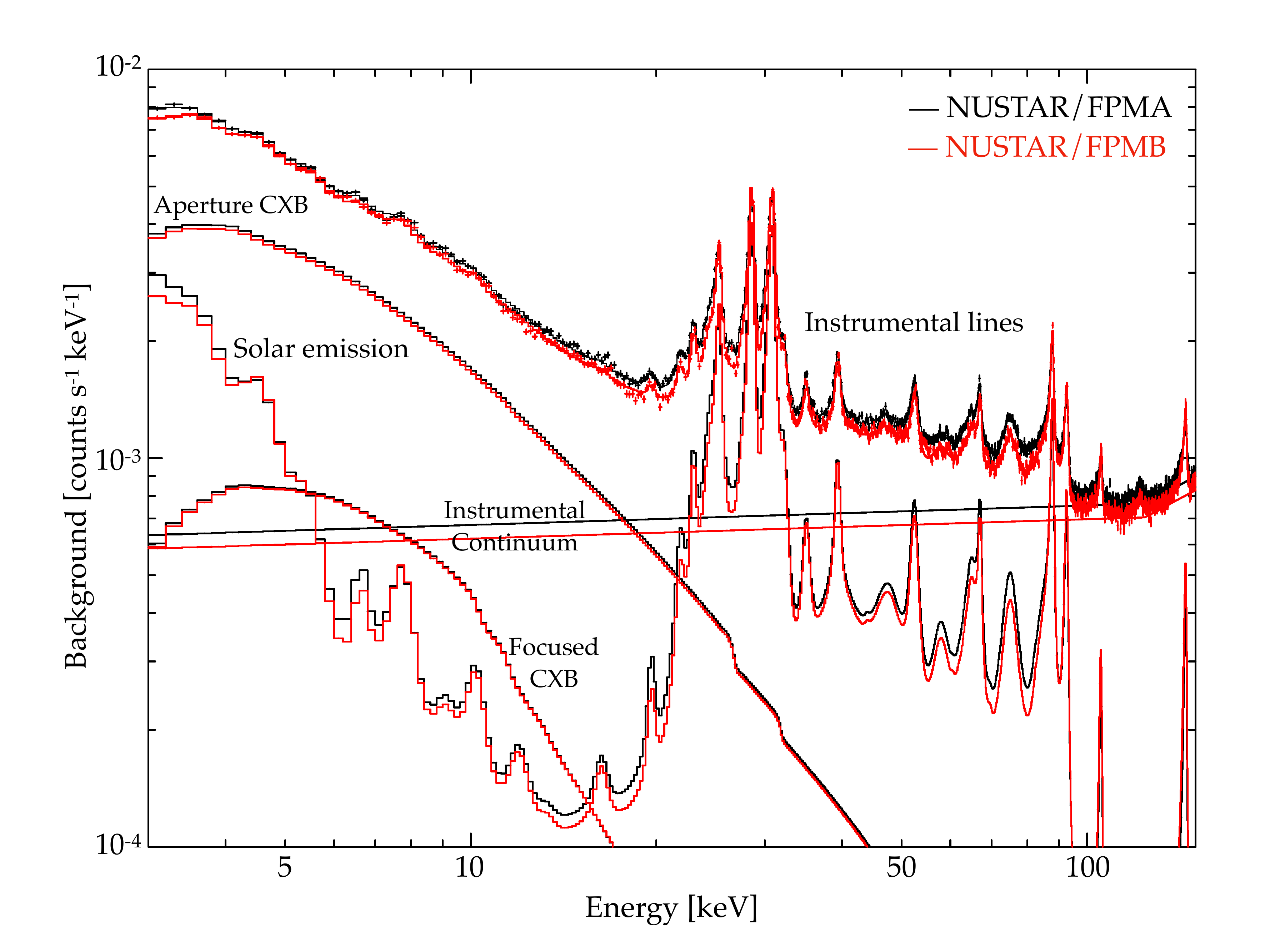}
\caption{Example of activation-induced background lines. The plot shows the major contributions to the overall background measured by the two instruments onboard the NuSTAR observatory (e.g., solar particles, aperture CXB and so on). The contribution of activation-induced lines is well apparent in the 20--200 keV range. Adapted from \cite{wik14}. }
\label{f:wik14}
\end{figure}

Besides the short and long-term background count rate effects due to the activation, the decay of particular isotopes can produce particular X and gamma-ray lines in the measured spectra. Likewise, details of line energy and intensity will depend on the activated materials and irradiation profile. Figure~\ref{f:wik14} shows, as an example, the X-ray background measured by the NuSTAR observatory: several activation (and fluorescence) induced lines are prominent in the 20--200 keV range.

\section{Background simulation, mitigation and evaluation strategies}\label{s:montecarlo}

\subsection{The Monte Carlo approach}
Estimating the level of on-board background for a X-ray space experiment is a mandatory feature of the instrument design phase. Given the expected and modelled environment, various approaches for the verification of the compliance with the mission scientific requirement and for the background minimisation could be adopted (e.g., hardware shielding, on-board software filtering or mission-wide strategies such as an optimised pointing plan).

In this aspect, usually a Monte Carlo simulation is performed, involving radiation-transport codes which are able to follow any possible interaction of a photon or a particle with the spacecraft and detector structures, each described with its geometry and composing materials (see, e.g., Figure~\ref{f:exampleMC}). The most common framework used for this regard is Geant-4 \cite{agostinelli03} usually adopted as a standard for high-energy physics experiments. Geant-4 is a collection of C++ libraries which can be employed to build a ``virtual'' model (\emph{mass model}) of the instrument, including a geometrical and physical description, and to handle the simulation of the possible physical interactions of the environment particles with the instrument, extracting the meaningful outcomes such as the energy deposits inside the detector.

The computational resources required to run a simulation will depend on the desired statistics, but also on the level of detail of the mass model. For this reason, usually the geometrical description is implemented in a manner that is progressively less accurate the further from the most critical or sensitive parts of the detector. The spacecraft structures, for example, can be designed as rough geometrical ``dummies'', with an approximate shape and chemical composition but realistic volumes and masses, while the detector itself can be a CAD-derived, detail-full design.
Also the level of simulation accuracy (the ``range cut'' of the physical processes implemented) can be defined on a custom basis.

\subsection{Mitigation strategies}

Depending on the detector architecture, events due to charged particles be recognised and filtered out, on the basis of their properties (such as the energy or the number of triggered pixels for a segmented or imaging detector).

\begin{figure}[htbp]
\centering
\includegraphics[width=0.8\textwidth]{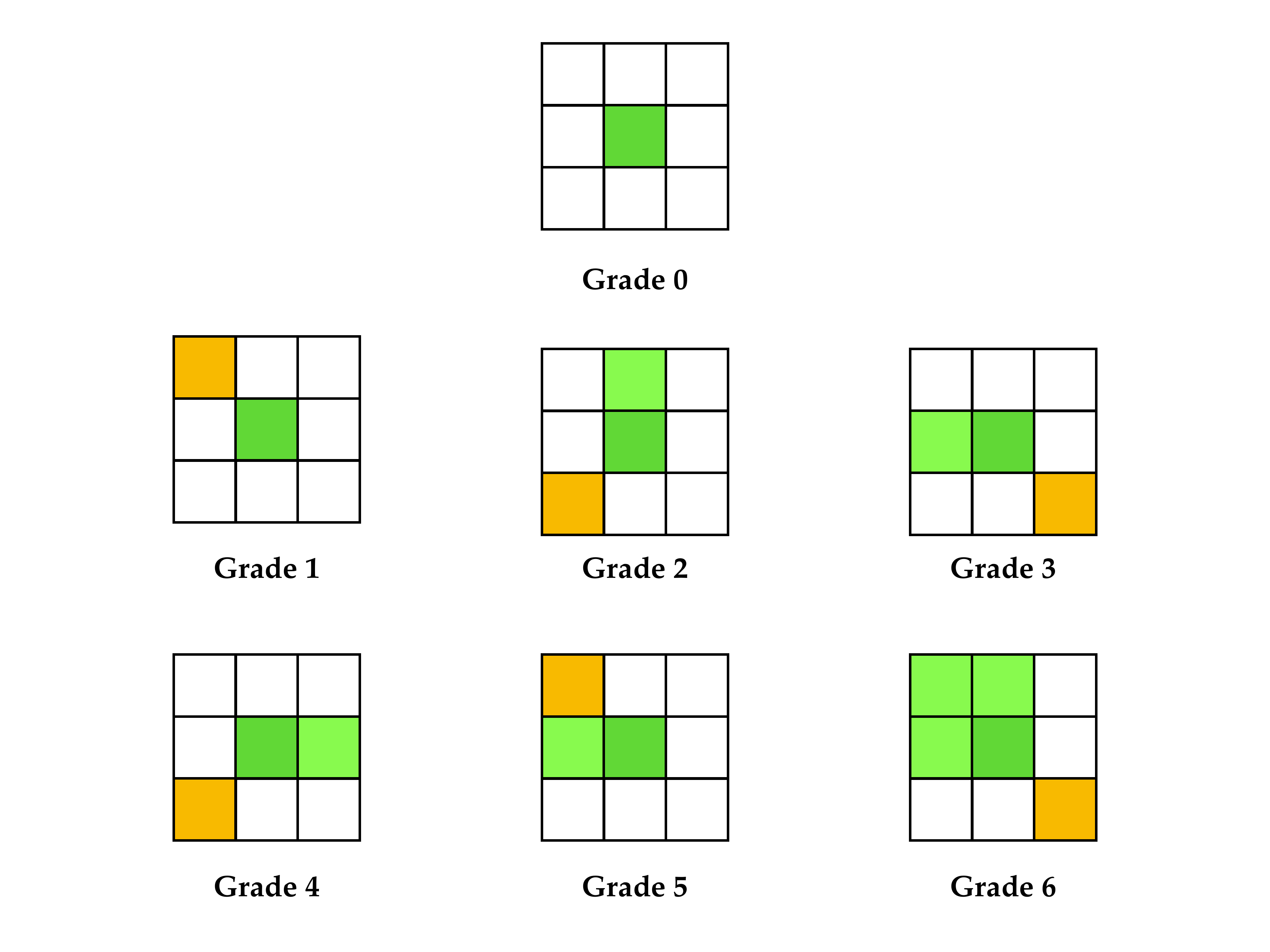}
\caption{Example of event grade filter. The definitions of the seven ASCA/SIS patterns are shown: the central pixel is the one with the highest signal. ``Split-charge'' pixels are in different shades of green, while ``detached corner'' pixels in orange. Event of grade 1, in which a central pixel is accompanied by a detached corner pixel with a certain amount of charge, are usually associated with particle events and thus filtered out.}
\label{f:eventgrade}
\end{figure}

In particular,  an approach usually adopted for CCD-based detectors is to identify the \emph{event grade}, i.e., the pattern of charge on pixels around a given one with the highest pulse (Figure~\ref{f:eventgrade} shows an example). Usually, the pattern in a pixel detector is distinct between events due to a photon to a particle, and this applies even for particles which enter perpendicularly to the detection plane.

\begin{figure}[htbp]
\centering
\includegraphics[width=0.8\textwidth]{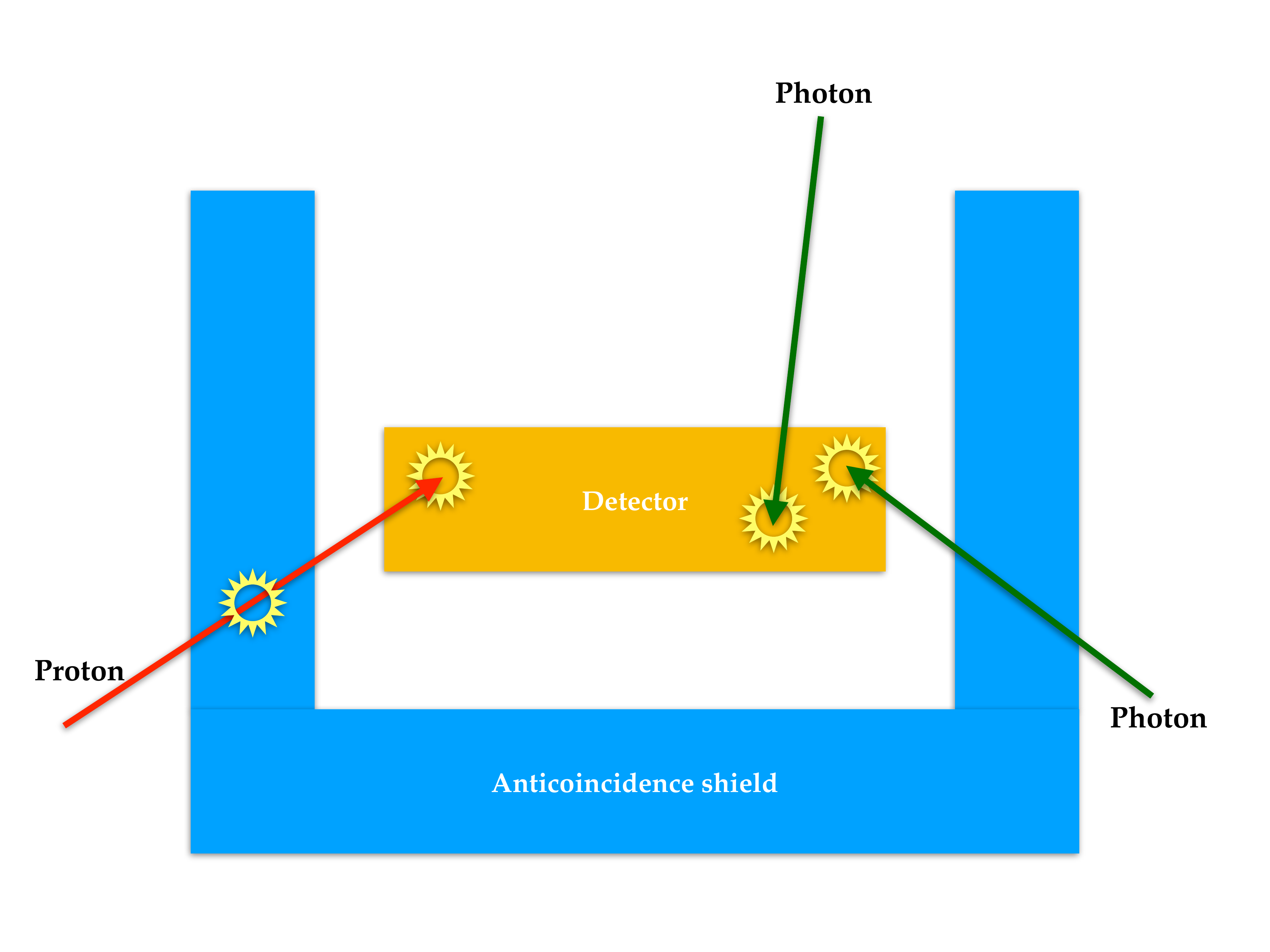}
\caption{Conceptual example of an active anticoincidence. A charged particle will leave a signal on both the detector and on the external shield, while photons will interact only with the detector. A suitable control electronics can then veto events occurring on both structures.}
\label{f:anticoincidence}
\end{figure}

Another technique to suppress the contribution of charged particles to the instrumental background is to use an \emph{active anticoincidence} (Figure~\ref{f:anticoincidence}), i.e., surrounding the instrument with a fast, particle-sensitive detector. Such an instrument can be, for example, a thin slab of plastic scintillator read out by a photosensor. The triggering of this anticoincidence detector will provide a veto signal allowing to discard any event occurring in the detector of interest by the same particle.

\begin{figure}[htbp]
\centering
\includegraphics[width=0.75\textwidth]{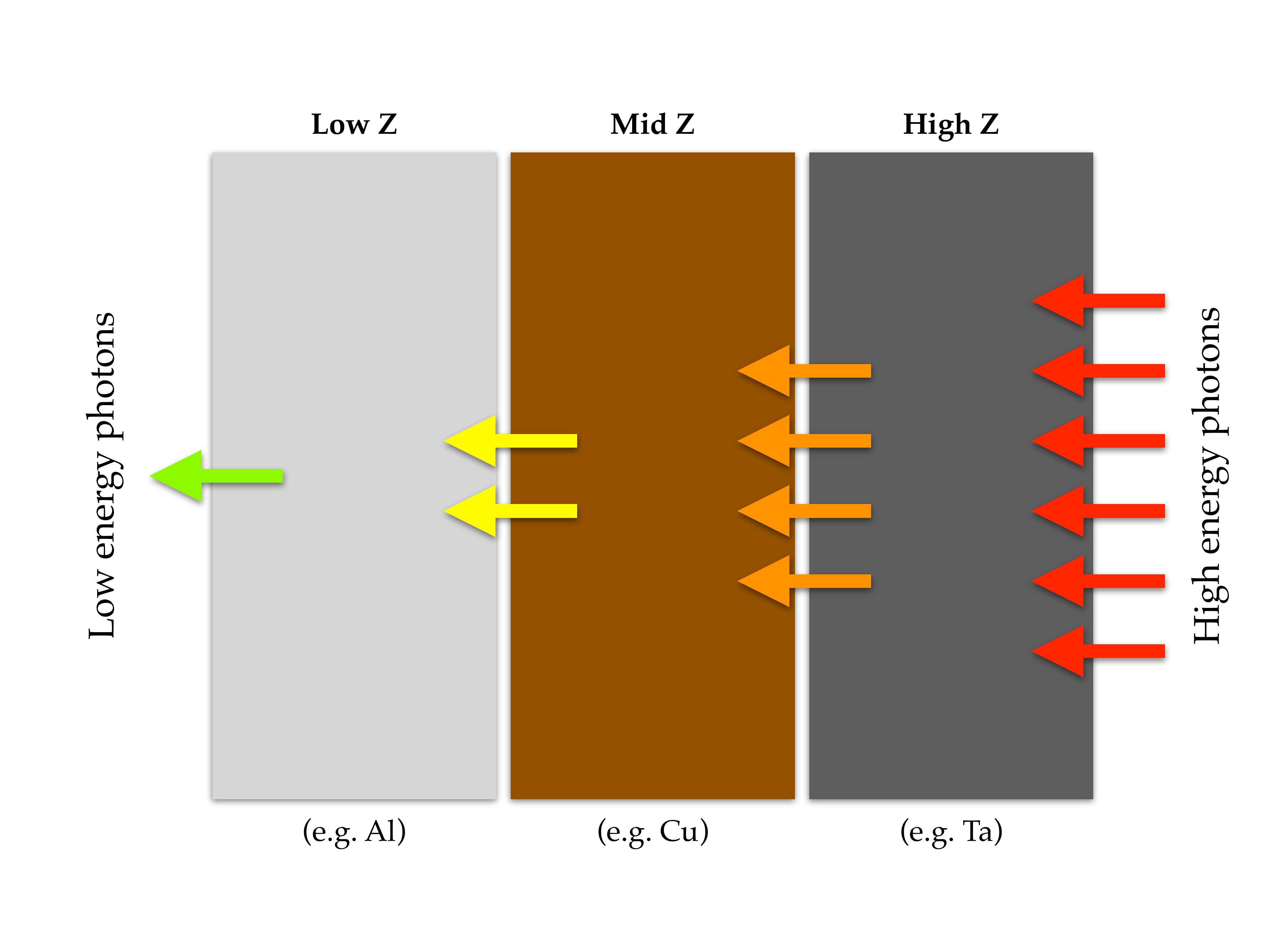}
\caption{Example of a graded-$Z$ shield, in which layers of materials with different atomic number progressively filter the incoming photon flux, shifting the typical energies to lower values.}
\label{f:gradedZshield}
\end{figure}

To suppress the background due to high-energy photons, \emph{shielding} of the detectors using suitable thicknesses of a high-$Z$, high density material can be employed. However, these shields can still produce a large amount of secondary radiation, e.g. due to fluorescence. A more refined technique, which also allows to save mass, is to use a \emph{graded-Z shielding} (Figure~\ref{f:gradedZshield}), a sandwich of layers of materials with increasingly lower atomic number.
The layer with the highest $Z$ can scatter and absorb protons, electrons and $\gamma$-rays. Each subsequent layer absorbs the X-ray fluorescence produced by the interactions of  photons in the previous material, eventually lowering the energy to a suitable level.

\subsection{On-board or on-ground evaluation}

\begin{figure}[htbp]
\centering
\includegraphics[width=0.5\textwidth]{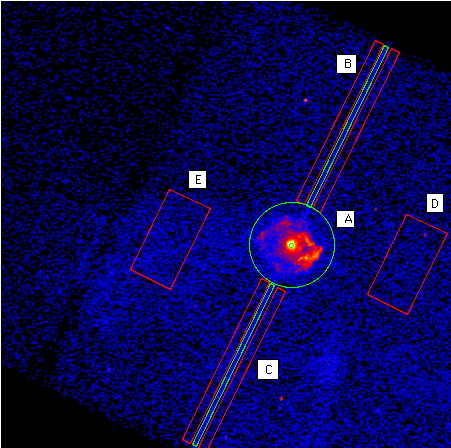}
\caption{Example of background selection on image data. The figure shows a \emph{Chandra} observation of the pulsar wind nebula (PWN) around the pulsar PSR~B0540-69 in the Large Magellanic Cloud. The ``source'' regions, encircling the PWN, are outlined in green, while the orange boxes delimit the ``background'' extraction regions.}
\label{f:0540}
\end{figure}

For an imaging instrument, the usual approach is to perform a selection by filtering out the detector pixels directly illuminated by the source (usually within an extraction radius comparable with the instrumental point spread function, PSF) and selecting a nearby image region which is expected to be affected only by the background. The data from the latter region will be conveniently scaled in order to evaluate the background contribution to the emission collected within the source region (Figure~\ref{f:0540}). This approach becomes more complex when observing extended or large sources, in which the background extracted far from the region of interest could not be representative of the actual one (e.g., for vignetting due to the optics, or for detector location-dependent effects).

\begin{figure}[htbp]
\centering
\includegraphics[width=\textwidth]{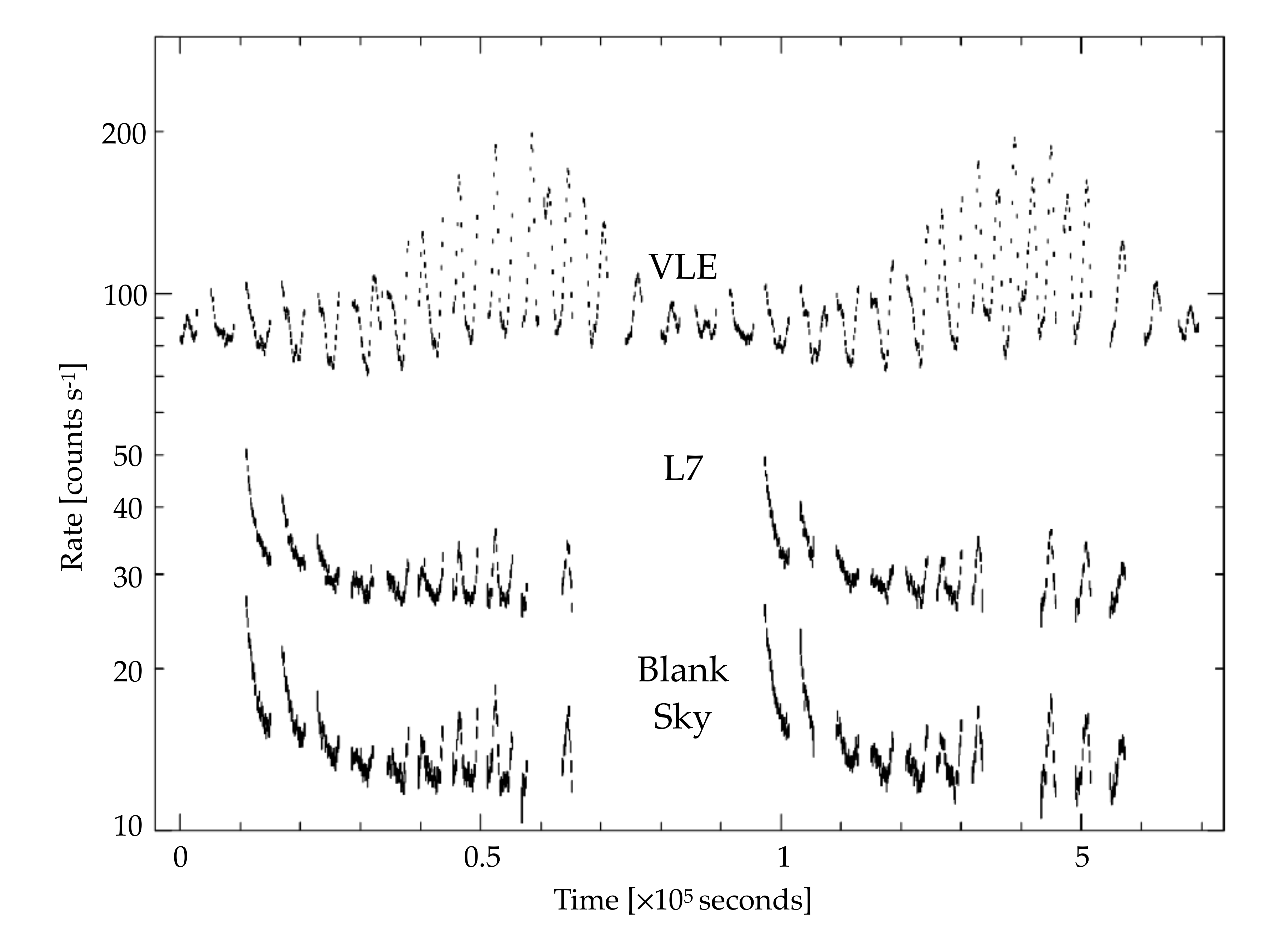}
\caption{Count rates from a blank-sky observation performed with one unit of the PCA onboard RXTE. 
The orbital variations, the effect of SAA passages and observational gaps are easily apparent. The VLE and L7 instrumental rates are shown to be highly correlated with the blank-sky (background) count rate and thus used in the parameterised model of the instrumental background during an observation. Adapted from \cite{jahoda06}.}
\label{f:rxtebkg}
\end{figure}

Several other approaches are possible. One is is to \emph{model} the background, using dedicated blank-sky observations or using other measurements as a proxy. For example, in the PCA (Proportional Counter Array) instrument onboard the Rossi X-ray Timing Explorer (RXTE) mission, two housekeeping rates from the proportional counters used in that instrument (i.e., the pairwise coincidence rate and very large, saturating, event rate, see \citep{jahoda06} for details) were found to be extremely well correlated with the background count rate measured during observation of source-free (``blank-sky'') regions (Figure~\ref{f:rxtebkg}). As a consequence, a model of the instantaneous background rate was constructed using these instrumental housekeepings as one of the input parameters, beside other quantities such as the epoch or the elapsed time from the last SAA passage.

For a non-imaging instrument, another approach is to alternate between observations of a sky region containing the source and of a nearby, hopefully source-free region containing only background.

\section{Summary and conclusions}
In-orbit background is a feature of every X-ray detector operating in orbit. As we have seen, according to the detector type, instrument design, mission architecture and operating orbit the details of the background properties and their impact on the scientific observations can greatly vary. 

The design of a new X-ray mission will unavoidably involve an estimation of the background, taking into account the expected scientific outcomes and the resulting sensitivity level. Monte Carlo simulations of the detector architecture is an essential tool in an iterative framework, in which an increasing more accurate detector description and background source modelling will interplay with a careful study of the possible design trade-offs aiming to minimise the overall background impact on the observations, and maximise the detector performance.

% BibTeX users please use
%\bibliographystyle{spphys.bst}
\bibliography{bibliography.bib}

\end{document}